\documentclass[fleqn,usenatbib]{mnras}

\usepackage{newtxtext,newtxmath}

\usepackage[T1]{fontenc}
\usepackage{ae,aecompl}

\usepackage{graphicx}	
\usepackage{amsmath}	
\usepackage{amssymb}	
\usepackage[normalem]{ulem} 

\newcommand{\astr}{\emph{AstroSat}}
\newcommand{\nus}{\emph{NuSTAR}}
\newcommand{\src}{MAXI~J1535$-$571}

\newcommand{\yb}[1]{{\color{black}#1}}

\title[Spectro-timing analysis of \src]{Spectro-timing analysis of \src\ using \astr}

\author[Bhargava et al.]{
Yash Bhargava,$^{1}$\thanks{E-mail: yash@iucaa.in} 
Tomaso Belloni,$^{2}$
Dipankar Bhattacharya,$^{1}$
and Ranjeev Misra,$^{1}$
\\
$^{1}$Inter-University Centre for Astronomy and Astrophysics Post Box No. 4, Ganeshkhind, Pune-411007, India\\
$^{2}$INAF, Osservatorio Astronomico di Brera, Via E. Bianchi 46, I-23807 Merate (LC), Italy
}

\date{Accepted XXX. Received YYY; in original form ZZZ}

\begin{document}
\label{firstpage}
\pagerange{\pageref{firstpage}--\pageref{lastpage}}
\maketitle

\begin{abstract}
We report the results of the analysis of an \astr\ observation of the Black Hole candidate \src\ during its Hard Intermediate state. We studied the evolution of the spectral and timing parameters of the source during the observation. The observation covered a period of $\sim$5 days and consisted of 66 continuous segments, corresponding to individual spacecraft orbits. Each segment was analysed independently. The source count rate increased roughly linearly by $\sim$30\%. We modelled the spectra as a combination of radiation from a thermal disk component and a power-law. The timing analysis revealed the presence of strong Quasi Periodic Oscillations with centroid frequency $\nu_{\rm{QPO}}$ fluctuating in the range 1.7--3.0~Hz. We found a tight correlation between the QPO centroid frequency $\nu_{\rm{QPO}}$ and the power-law spectral index $\Gamma$, while $\nu_{\rm{QPO}}$ appeared not to be correlated with the linearly-increasing flux itself. We discuss the implications of these results on physical models of accretion. 

\end{abstract}




\section{Introduction}

Black-Hole Binaries (BHBs) are stellar systems in which one of the objects is a stellar-mass black hole and the companion star is typically a low-mass star that fills its Roche lobe, leading to accretion of matter onto the black hole, or a high-mass star feeding the black hole through its stellar wind. The majority of these systems are transient, with only a few known persistent sources. The evolution of their properties, in particular during transient outbursts when the accretion rate swing is large, is characterised by a series of source states, defined through the spectral and fast-variability properties   \citep[see][]{Done2007A&ARv..15....1D, Belloni2010LNP...794...53B, Belloni2011BASI...39..409B}. 

The black hole binaries typically follow a hysteresis loop in the Hardness-Intensity diagram and different positions on this diagram correspond to different states of the system. The Low Hard state (LHS) of the system is characterised by a hard spectrum and high fractional rms variability \citep[$\sim$30\%,][]{Belloni2005AIPC..797..197B}. As the source evolves into a Hard-Intermediate State (HIMS) the spectrum of the source softens and indicates the presence of a thermalised disk. Low frequency Quasi Periodic Oscillations (QPOs) of type C are also detected in these states \citep{Casella2005ApJ...629..403C}. The source can then evolve into a Soft-Intermediate state (SIMS) which is characterised by a softer spectrum with the disk component dominating the flux. Transient QPOs of Type A and B are also seen in the power spectrum from this state \citep{Casella2005ApJ...629..403C}. The source then typically evolves into a High Soft state (HSS) in which the spectrum is strongly dominated by a thermalised disk. The power spectrum can be fitted with a flat power law. The source typically fades and returns to the LHS. 

QPOs are ubiquitous features in the variability pattern of BH binary systems \citep[see][and references therein]{Belloni2014SSRv..183...43B}. Low frequency QPOs (0.1--30~Hz) are associated with oscillations in the inner regions of the accretion disk. They are observed in the hard states, LHS and HIMS. Their energy spectrum indicates that their origin is connected to the high-energy component and not the thermal disk component. As their frequency is too low to be directly associated to Keplerian motion in the inner region of the accretion flow, models have concentrated on other physical time scales. The RPM model \citep{Stella1998ApJ...492L..59S, Stella1999ApJ...524L..63S} associates these oscillations to the Lense-Thirring precession frequency at a certain radius of the accretion flow. A more complex model connected to accretion has been proposed, which takes into account a broader precessing region surrounded by a thermal disk \citep[][and references therein]{Ingram2011MNRAS.415.2323I,Ingram2016AN....337..385I}, as in the truncated-disk paradigm \citep[see][]{Done2007A&ARv..15....1D}.

The study of the correlation between spectral and timing properties can help in constraining the theoretical models of accretion disks around a BH. In particular the dependence of the QPO centroid frequency on spectral parameters is a crucial observable. 
A deep study of the correlation between QPO centroid frequency ($\nu_{\rm{QPO}}$) and slope of the high-energy power law ($\Gamma$) for the peculiar system GRS 1915+105 was conducted by \citet{Vignarca2003A&A...397..729V}. In this work the authors extract and present this correlation also for other systems (GRO J1655-40, XTE J1550-564, XTE J1748-288 and 4U 1630-47), showing that it is a general property for BH binaries. The correlation is a positive one: higher QPO frequencies are associated to steeper energy spectra and a turnoff at high frequencies is observed. This work and subsequent works by \cite{Shaposhnikov2007ApJ...663..445S} and \cite{Shaposhnikov2009ApJ...699..453S} sample two different regimes of variations in the parameters: for GRS 1915+105 fast variations, within an hour, are considered, while for the other more conventional transient systems like Cygnus~X-1 the values come from observations spread throughout an outburst, i.e. months.

\src\ was \yb{discovered} independently by MAXI and \textit{Swift}-BAT on September 02, 2017 \citep{Negoro2017ATel10699....1N, Markwardt2017GCN.21788....1M}. 
\yb{ \cite{Kennea2017ATel10700....1K} provided a more accurate position of the source using \textit{Swift}-UVOT and \textit{Swift}-XRT observations.}
\citet{Scaringi2017ATel10702....1S} reported the detection of an optical counterpart to \src, followed by the detection of near infrared radiation by \citet{Dincer2017ATel10716....1D}. Radio detection from \emph{ATCA} \citep{Russell2017ATel10711....1R} \yb{and further brightening in the X-ray flux \citep{Negoro2017ATel10708....1N}} suggested the nature of the compact object as a black hole. \citet{Mereminskiy2017ATel10734....1M} detected low frequency QPOs at 1.9~Hz in Swift XRT observations conducted on September 11, 2018, indicating that the source was in the Hard-Intermediate state, i.e. in transition from the hard to the soft state. \yb{\citet{Mereminskiy2018AstL...44..378M} and \citet{Stiele2018ApJ...868...71S} discuss the evolution of the QPO frequency and observe a positive correlation between the QPO centroid frequency and the power law index of the spectrum, relating the QPO centroid frequency to the inner truncation radius of the disk, \citet{Mereminskiy2018AstL...44..378M} derive self-consistent results for the physical parameters of the inner Comptonising cloud.} 
\citet{Xu2018ApJ...852L..34X} analyse the \nus\ spectrum and constrain the spin of the black hole to $>0.84$ and inner truncation radius to $<2.01$~R$_{\rm{ISCO}}$. \yb{\citet{Sreehari2019MNRAS.tmp.1270H} and \citet{Sridhar2019arXiv190509253S} analyse the spectral properties of the source using the data from the \astr\ observation and constrain the mass of the source to 5.14--7.83~M$_{\odot}$ and $10.39^{+0.61}_{-0.62}$~M$_{\odot}$ respectively. \citet{Sreehari2019MNRAS.tmp.1270H} also present the evolution of the timing parameters using \emph{Swift}-XRT and LAXPC observations and classify the states of the source using the QPOs detected in in the power spectra. }

In this paper, we report the result of the spectral-timing analysis of the \astr\ SXT+LAXPC data of \src\ obtained over a period of five days and concentrate on the $\nu_{\rm{QPO}}$-$\Gamma$ correlation.

\section{Observation and Data analysis}\label{sec:obs}

\begin{figure}
\centering
\includegraphics[width=0.8\columnwidth]{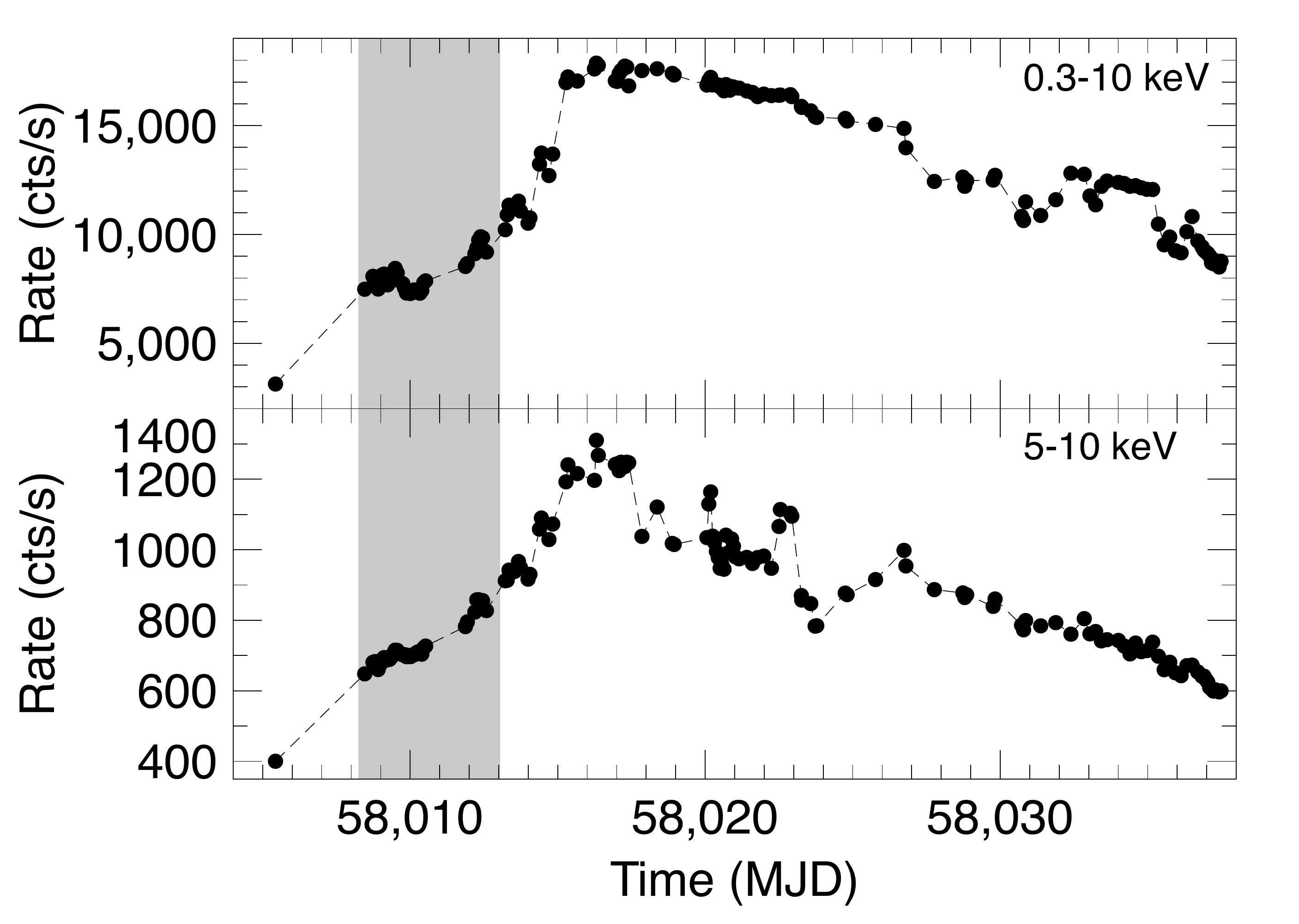}
\caption{NICER light curves of \src\ (top panel: 0.3--10 keV, bottom panel: 5--10 keV) over the period 2017 Sep. 9 to Oct. 11, with the time interval of the \astr\ observation marked in grey.}
\label{fig:nicer_licu}
\end{figure}

\begin{figure}
\centering
\includegraphics[width=0.8\columnwidth]{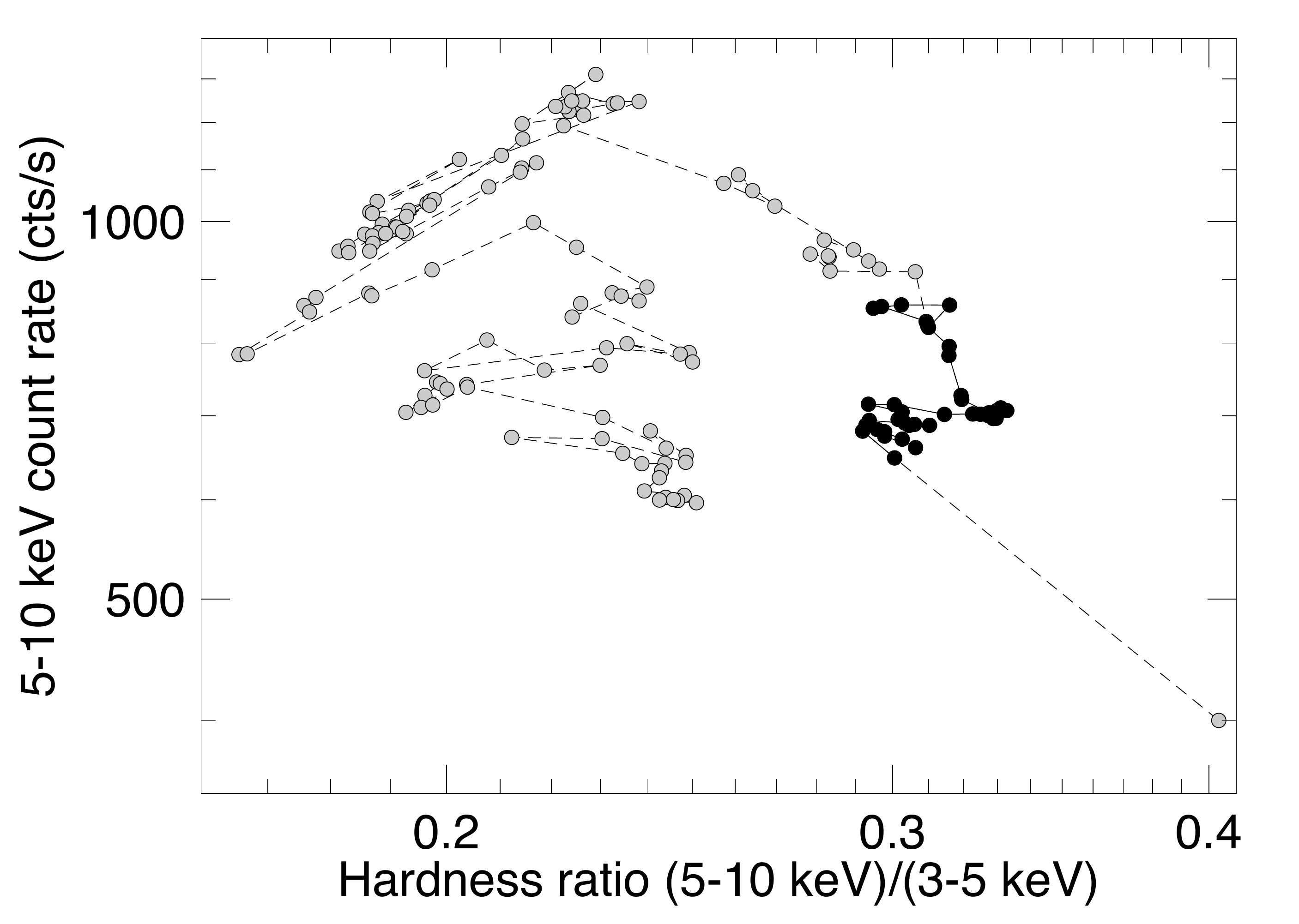}
\caption{Hardness-intensity diagram of \src\ as observed by NICER over the same interval as \autoref{fig:nicer_licu}. The energy bands for the two axes are indicated in the label. The black points are those within the observation window of \astr. The dashed lines mark a longer time gap (see \autoref{fig:nicer_licu}).}
\label{fig:nicer_hid}
\end{figure}

Based on a trigger from \cite{Negoro2017ATel10699....1N}, \astr\ \citep{Singh2014SPIE.9144E..1SS} triggered a Target of Opportunity (ToO) observation (Observation ID: 01536) of \src\ from September 12, 2017 (MJD 58008.2309) to September 17 2017 (MJD 58013.1545). The source was observed in the rising part of the outburst. In order to place the \astr\ data in the context of the outburst, we analyzed data from the Neutron star Interior Composition Explorer (NICER) mission on board the International Space Station \citep{Gendreau2017AAS...22930903G..NICER}, which observed the target regularly over the period September 9 to October 11 2017 and provided a good picture of the overall evolution of the first part of the outburst. We extracted the NICER count rates from all detectors without background subtraction since the source is very bright and produced light curves in different energy bands as well as hardness ratios. In \autoref{fig:nicer_licu}, we show the NICER light curves in the full 0.3--10 keV energy band (top panel) and in the 5--10 keV energy band, closest to the LAXPC coverage. The grey band in the figure represents the time interval of the \astr\ observation. It is clear that \astr\ observed during an interval of roughly monotonic rise of flux in the 5--10 keV band, while the full NICER light curve is more complex and in particular shows a faster flux increase after the \astr\ coverage.
The Hardness-Intensity Diagram (HID) from the NICER data, a useful tool to represent the evolution of a black-hole transient \citep[see e.g.][]{Belloni2011BASI...39..409B} is shown in \autoref{fig:nicer_hid}, where the points covering the \astr\ observation period are marked in black. The typical counter-clockwise evolution can be seen \citep[][]{Belloni2011BASI...39..409B}, but in order to fully classify the source states additional timing analysis needs to be done (see below).

\yb{Primary instruments onboard \astr\ which observed the outburst include the Soft X-ray Telescope (SXT) and the Large Area X-ray Proportional Counter (LAXPC). SXT is a focussing X-ray telescope which operates in the energy range of 0.3--8~keV with an energy resolution of 5--6\% at 1.5~keV and an effective area of $\sim$128~cm$^2$ at 1.5~keV \citep{Singh2017JApA...38...29S}. The observation of the source was carried out with SXT operating in Fast Window timing (FW) mode. This mode observes only the central 150$\times$150 pixels from the total 600$\times$600 pixels. Due to the large Point Spread Function (PSF) of SXT, the source occupied the complete field of view (FOV; 10\arcmin) in FW mode. However the smaller number of pixels to be read out allowed a better time resolution (0.278~s) than the full frame readout (2.37~s)} 

LAXPC is an X-ray proportional counter array operating in the range 3--80~keV with an energy resolution of 10--15\% at 20~keV. The timing resolution of the instrument is 10~$\mu$s with a dead time of 42~$\mu$s. There are three identical detectors (referred to as LXP10, LXP20 and LXP30 respectively) on \astr\ with a combined effective area of 6000~cm$^2$ \citep{Yadav2016SPIE.9905E..1DY, Antia2017ApJS..231...10A}. LAXPC was operated in Event Analysis (EA) mode for the duration of the observation, which allowed for the data of individual photons to be available to the user.

\subsection{Data reduction}
\yb{The data reduction of the \astr\ observation was done using the instrument pipelines provided by the respective Payload Operation Centres (POC). SXT data was reduced using SXT pipeline \texttt{AS1SXTLevel2-1.4a} and the calibration files released with the pipeline. LAXPC data was reduced using the LAXPC pipeline \texttt{laxpc\_make\_event} from the package \texttt{laxpcsoft\_Sep12\_2017}. The package also includes the calibration files for all the units of LAXPC.}

To monitor the evolution of the source over the observation period we divided the observation into the 66 \astr\ orbits, which are separated by gaps due to Earth occultations and passage through the South Atlantic Anomaly (SAA). This resulted in 66 data segments roughly equally spaced and with comparable exposures. The brightness of the source allowed us to analyse the spectral and timing properties for each of the segments as each segment had a large number of source photons. The segment boundaries are listed in \autoref{tab:time_stamps}. Standard Good time intervals (GTIs) were applied to each of the segments to remove the section corresponding to occultation of the source by Earth and near the SAA region. The GTI were created using the tool \texttt{laxpc\_make\_stdgti} provided by the LAXPC POC. Some intervals showed count-rate dips that were identified as instrumental and removed. The orbit-wise data from SXT were merged using the \texttt{sxtevtmerger} tool provided by the SXT POC. The individual segments were extracted using \texttt{xselect} (HEASoft version 6.23) on the merged event file and the GTI used for LAXPC. 
For each orbit of \astr\ the SXT has a lower exposure than LAXPC due to additional constraints (including reflection from the earth, larger SAA window). Due to the jitter in satellite pointing, the SXT coverage is also lower as the source occasionally moved out of the reduced window used in the Fast Window (FW) mode.

\subsection{Spectral analysis}\label{subsec:spec}
The spectra for all the segments were extracted using \texttt{xselect} for SXT and \texttt{laxpc\_make\_spectra} for LAXPC. As the observation was conducted with the SXT operating in FW mode, all the photons observed by SXT were assumed to be from the source. The spectra from the LAXPC was extracted for all instrument layers. 
However, due to a gas leak in LXP30, the response of the detector is considered uncertain and thus the data from this unit were not used for the spectral analysis, while they were retained for timing analysis.
For the SXT, as advised by the POC, standard response, ancillary response and background files were used for the analysis. In the case of the LAXPC, the responses were selected based on the spectral extraction and the background was modelled from the nearest blank-sky observation which had similar satellite position as the current observation.

The background estimates for both LXP10 and LXP20 are somewhat uncertain, so the spectra beyond 30~keV (where background starts dominating over the source) were ignored for the analysis. The response matrix modelling of SXT is uncertain below 0.8~keV due to the lack of a suitable calibration source. In our case, the source is brighter than the Crab and thus magnifies unmodelled instrumental features  below 1~keV. Therefore, the data below 1~keV were ignored. To use the $\chi^2$ statistics in judging the quality of the model to explain the data, we grouped the data from SXT such that each energy bin has at least 20 counts. The LAXPC has sufficient counts in each energy bin to allow for the $\chi^2$ statistics but the rebinning of the channels was done logarithmically, keeping dE/E at 5\% to account for the coarser energy resolution of the instrument. To account for the uncertainties in the response of the instruments a 3~\% systematic error was added to the model. 

Spectral analysis of \nus\ data of \src\ by \citet{Xu2018ApJ...852L..34X} indicates that the underlying compact object is a Black hole candidate. The authors have modelled the spectra by a thermal disk which is illuminated by a lamp post corona situated at $h=7.2^{+0.8}_{-2.0}\ r_{\rm{g}}$. The authors also report that the source has a particularly high absorption column density ($\rm{N_H}\sim8.2^{+0.3}_{-0.6}\times10^{22}$~cm$^{-2}$). This value is slightly higher than the value reported in the preliminary analyses by \citet{Kennea2017ATel10731....1K} (Swift/XRT; $\rm{N_H}\sim3.6\pm0.2 \times10^{22}$~cm$^{-2}$) and by \citet{Gendreau2017ATel10768....1G}  (NICER; $\rm{N_H}\sim4.89\pm0.06\times10^{22}$~cm$^{-2}$). \citet{Xu2018ApJ...852L..34X} have attributed this to the inclusion of the thermal disk in their modelling of the spectra. The observations from these different instruments were conducted at different epochs and thus the variation in the observed absorption could be due to internal changes in the source.  

\begin{figure}
    \centering
    
    \includegraphics[width=\columnwidth]{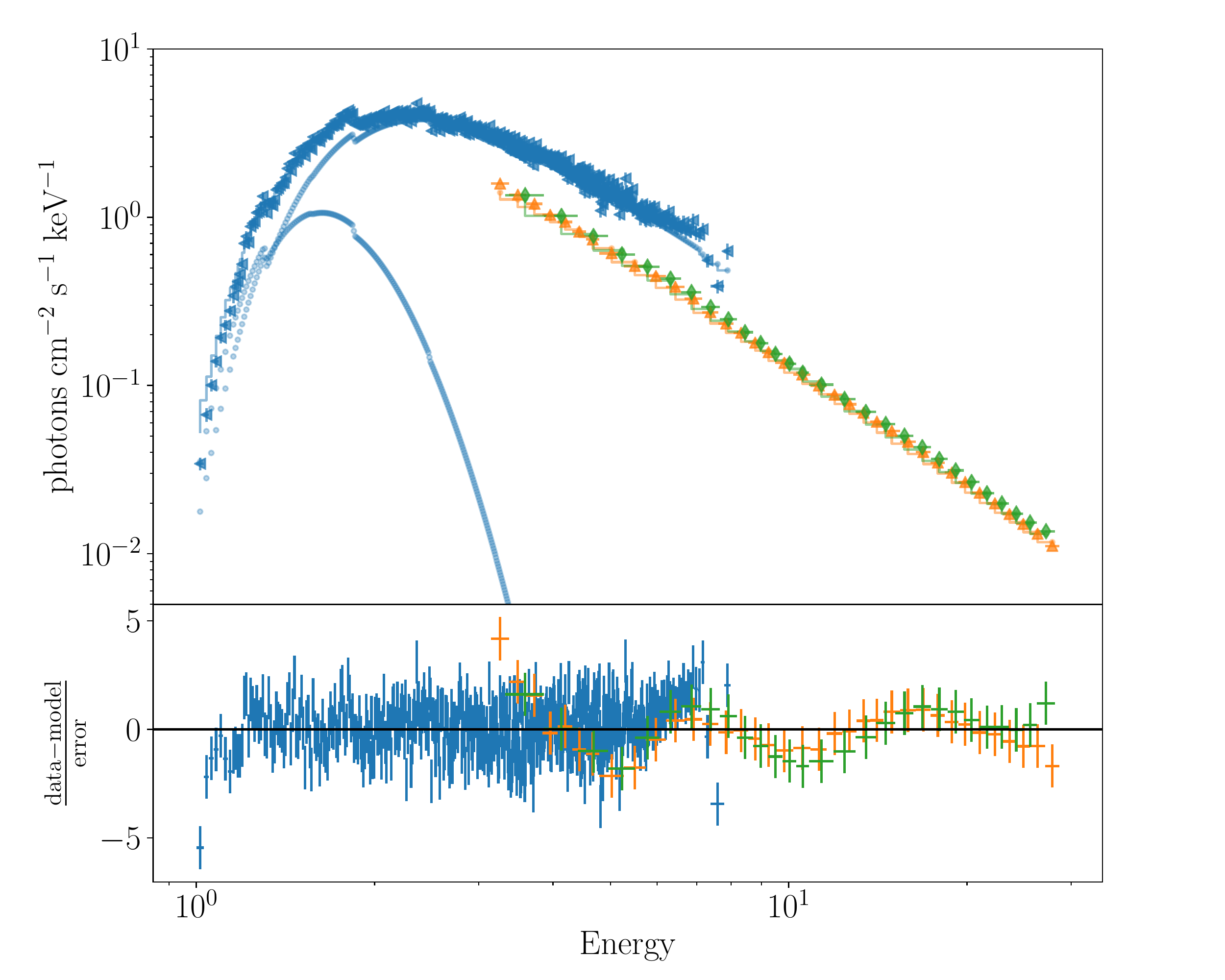}
    \caption{\yb{Typical energy spectrum of the source.    
    The blue left triangles, orange up triangles and green diamonds denote the spectrum observed by SXT, LXP10 and LXP20 respectively. 
    Spectrum from SXT covers the energy range 1--8~keV while both LAXPC units cover 3.5--30~keV. The top panel shows the unfolded spectrum. The slight difference in the level of the SXT and LAXPC spectra is known to be due to systematic errors in the model of SXT ARF. To indicate the shape of the spectra, the data has been highly rebinned. The colour version of the figure is available in online version of the manuscript.} }
    \label{fig:spec}
\end{figure}

\begin{figure}
\includegraphics[width=\columnwidth]{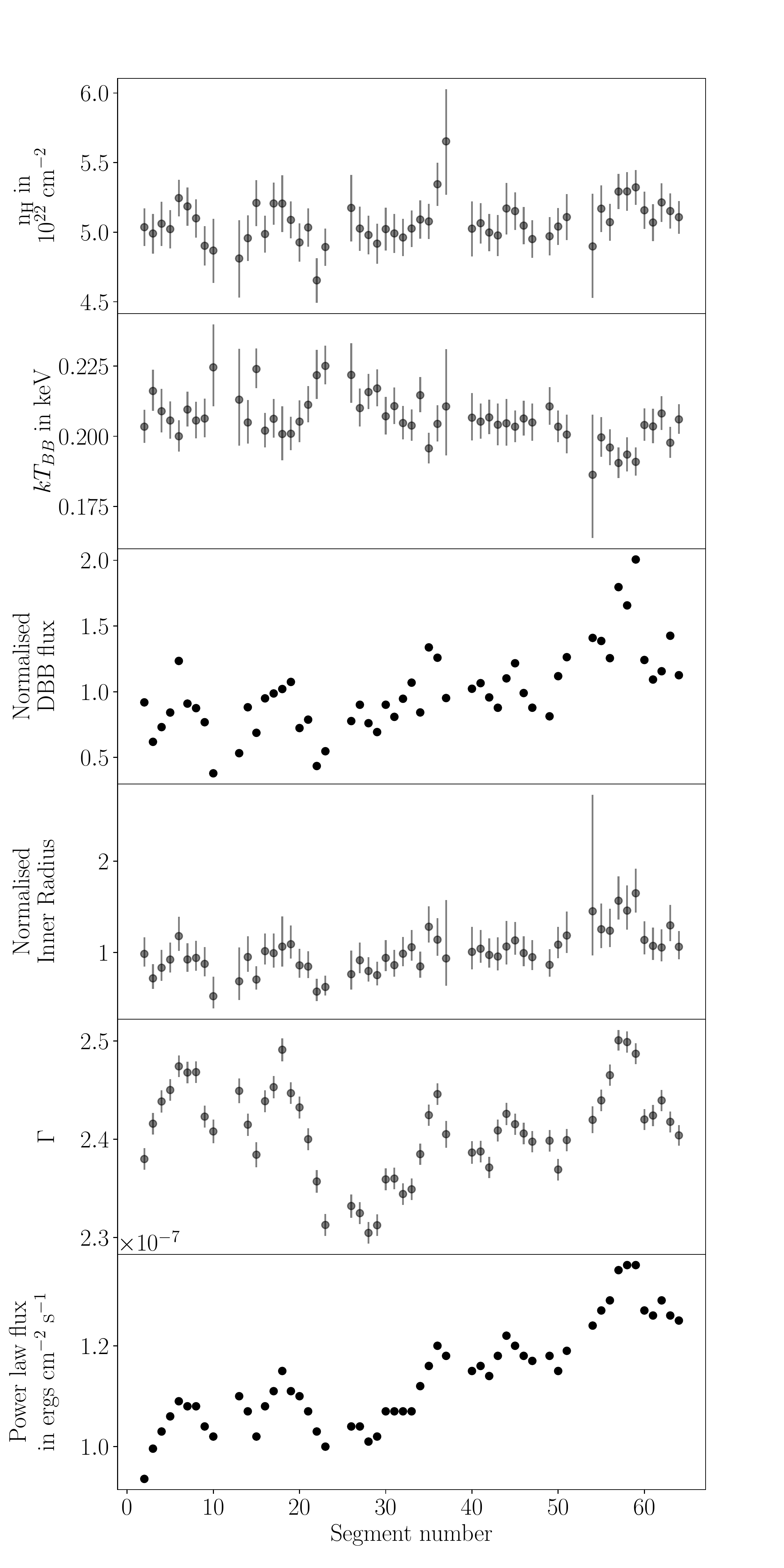}
\caption{Time evolution of the best-fit spectral parameters. In some of the segments, due to the satellite jitter, the source was outside the SXT field of view and thus no SXT spectra were available. The parameters for these segments could not be constrained well and thus they were excluded from further analysis and are not shown here. The disk flux is connected to the normalization of the component and is in arbitrary units.
}
\label{fig:para_var}
\end{figure}

In the present work, we modelled the source spectrum using a thermal disk and a power-law component. We did not find a significant presence of the Fe K$\alpha$ fluorescence line. \yb{Using the phenomenological model  of \citet{Xu2018ApJ...852L..34X} for the iron line and response matrices of SXT and LAXPC, we simulated a spectrum for the exposure of a typical segment. We find that the iron line was not required to have an acceptable fit to the simulated spectrum. Thus we claim that the iron line cannot be detected at a significant level for an exposure of the typical segment. We find a strong iron line when the spectra of all the segments are combined, the analysis of which will be reported in a future work.}
The absorption column density was modelled by \texttt{TbAbs} with \yb{the abundances from \citet{Wilms2000ApJ...542..914W} and the cross sections from \citet{Vern1996ApJ...465..487V}}. The model provided a satisfactory fit to the data with a typical reduced chi squared value of 1.1 for $\sim$700 degrees of freedom. A typical spectrum and the residuals to the fitted model is shown in \autoref{fig:spec}. 
The time evolution of the best-fit parameters is shown in \autoref{fig:para_var}. The absorption column density is around $5\times10^{22}$~cm$^{-2}$ for all the segments.  
\yb{\citet{Sreehari2019MNRAS.tmp.1270H} and \citet{Sridhar2019arXiv190509253S} report a slightly lower value for the same observation. The difference in the N$_{\rm{H}}$ is arising due to a difference in the ancillary response file (ARF) of SXT used by different authors. Our analysis uses the default on-axis ARF provided with the official pipeline.  }
The inner disk temperature stays roughly constant around the typical value of 0.2~keV. The power-law flux and the disk flux increase over the observation, but with significant fluctuations that appear to be correlated. Interestingly, the power-law index does not show a secular increase over the course of the observation.

\subsection{Timing analysis}\label{subsec:timing}

\begin{figure}
\includegraphics[width=\columnwidth]{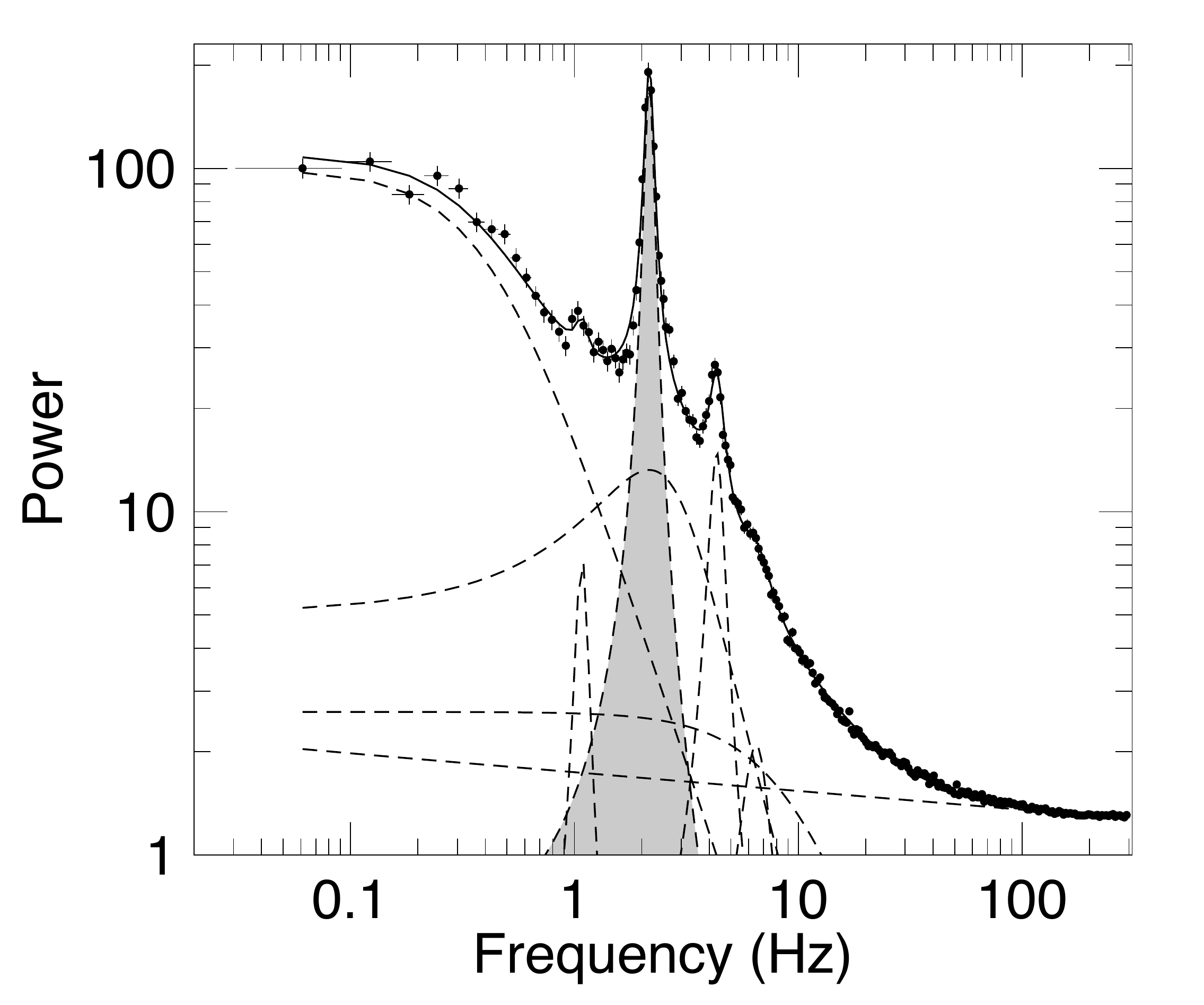} 
\caption{The Power Density Spectrum from the first segment. The thick line represents the best-fit model and the dashed lines the different components. The QPO peak at the fundamental frequency is marked in grey.}
\label{fig:pds}
\end{figure}

For each of the 66 segments, we produced power density spectra (PDS) from intervals of 16.384~s 
and averaged them. The final PDS were rebinned logarithmically before fitting. \yb{The PDS were extracted using the General High-energy Aperiodic Timing Software (GHATS version 1.1.1)\footnote{The software can be downloaded from  \url{http://www.brera.inaf.it/utenti/belloni/GHATS/Home.html}. The software is written by TMB. }}.  All the PDS display a clear QPO with harmonically related peaks. An example of a PDS is shown in \autoref{fig:pds}. The PDS were fitted with a model consisting of multiple Lorentzians: two flat-top components for the band-limited noise, up to four harmonically-related components for the QPO and its harmonics (sub-harmonic, fundamental, second harmonic and third harmonic, of which only the fundamental and the second harmonic are always detected) and a broad component for the L$_h$ feature \citep[see][]{Belloni2002ApJ...572..392B}. In addition, a flat power-law component was added to fit the Poissonian noise contribution. The fits were limited to the 0.0625--300 Hz range.

\begin{figure}
\includegraphics[width=\columnwidth]{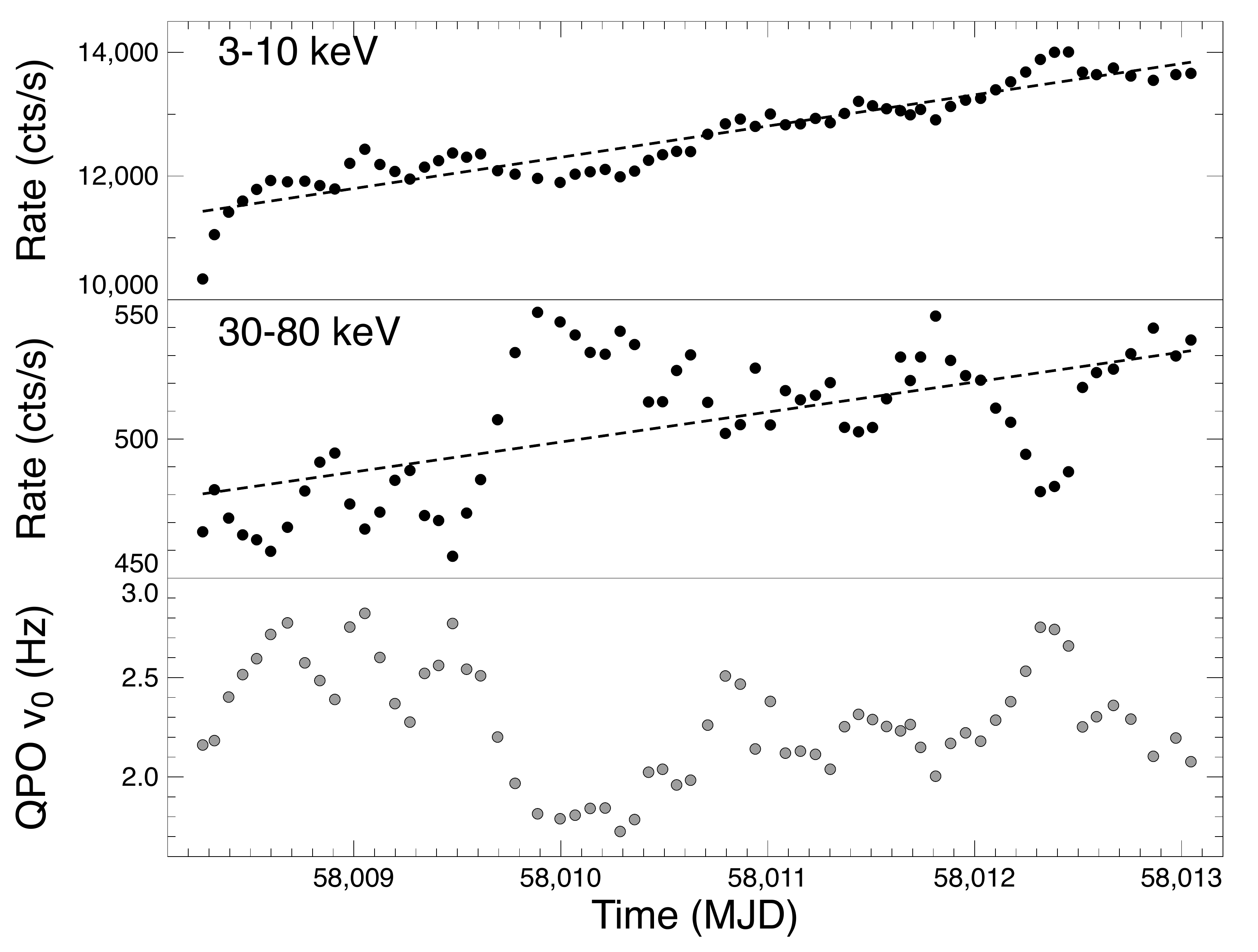} 
\caption{Top panel: LAXPC light curve in the 3--10 keV band with a linear fit. Middle panel:  light curve in the 30--80 keV band with a linear fit. Bottom panel: time evolution of the centroid frequency of the QPO. Errors are present in all panels, but are smaller than the symbols. 
}
\label{fig:time_evolution}
\end{figure}

Here we focus on the centroid frequency of the QPO. The signal is so strong that its measurement is largely model-independent. While the source count rate (and flux) increased roughly linearly throughout the observation, the QPO frequency varied in the 1.7--2.8~Hz range with no apparent correlation with the count rate. This can be seen clearly in \autoref{fig:time_evolution}. From the figure, one can see that the residuals from the linear fits to the two light curves are anti-correlated and 
 
\begin{figure}
\includegraphics[width=\columnwidth]{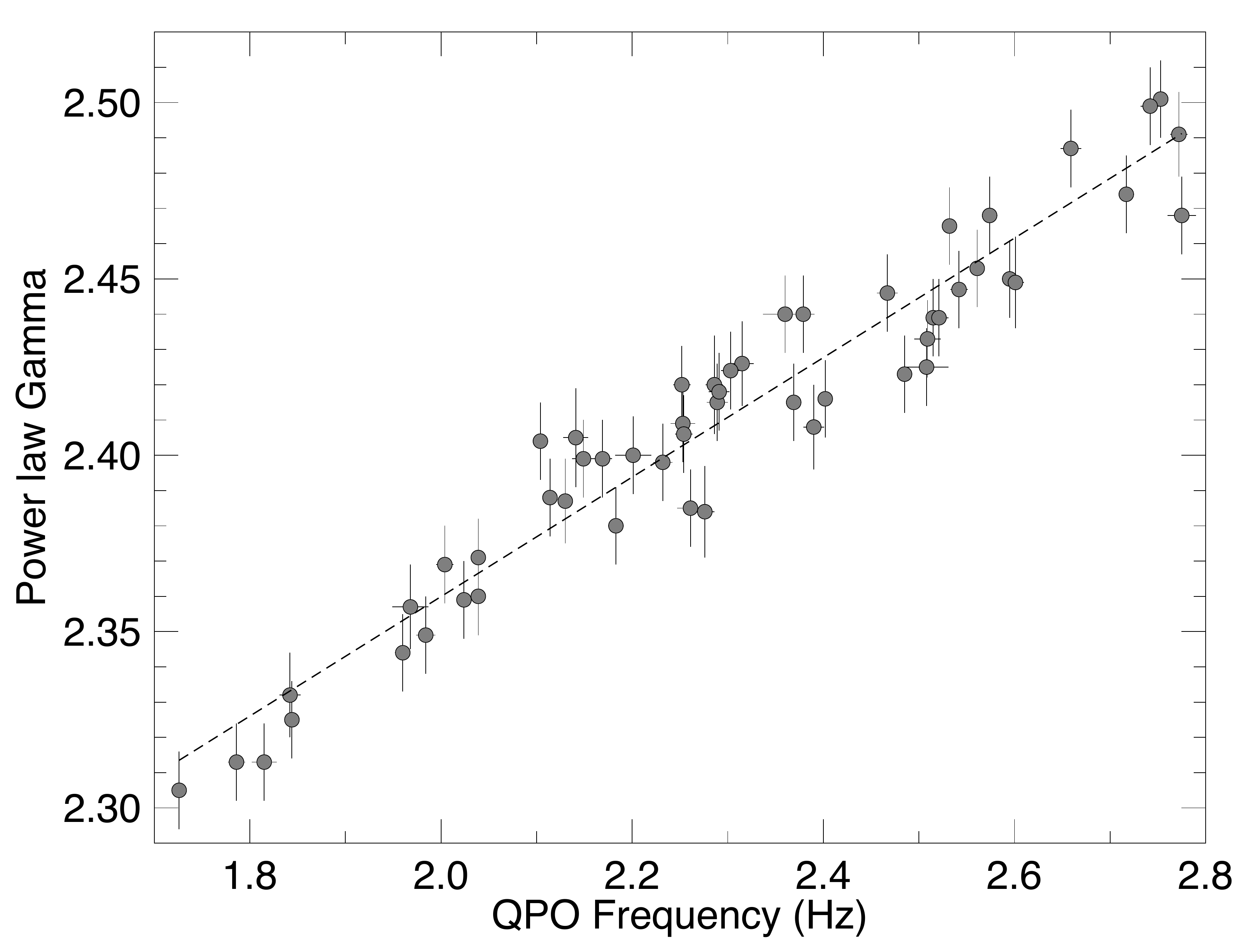} 
\caption{Correlation between power-law spectral index $\Gamma$ and QPO centroid frequency $\nu_{\rm{QPO}}$. The line is the best fit linear correlation. The segments in which SXT data were not available are not included.}
\label{fig:main_correlation}
\end{figure}

\yb{that $\nu_{\rm{QPO}}$ seems to have an anti-correlation with the residuals of the 30--80~keV countrate (and hence a possible positive correlation with those of 3--10~keV)}. Comparing the bottom panels of \autoref{fig:para_var} and \autoref{fig:time_evolution}, we can see that the QPO frequency is indeed well correlated with the power-law photon index as determined from the combined SXT+LAXPC spectral fits. The correlation is very tight and can be seen in \autoref{fig:main_correlation}.

\begin{figure}
\includegraphics[width=\columnwidth]{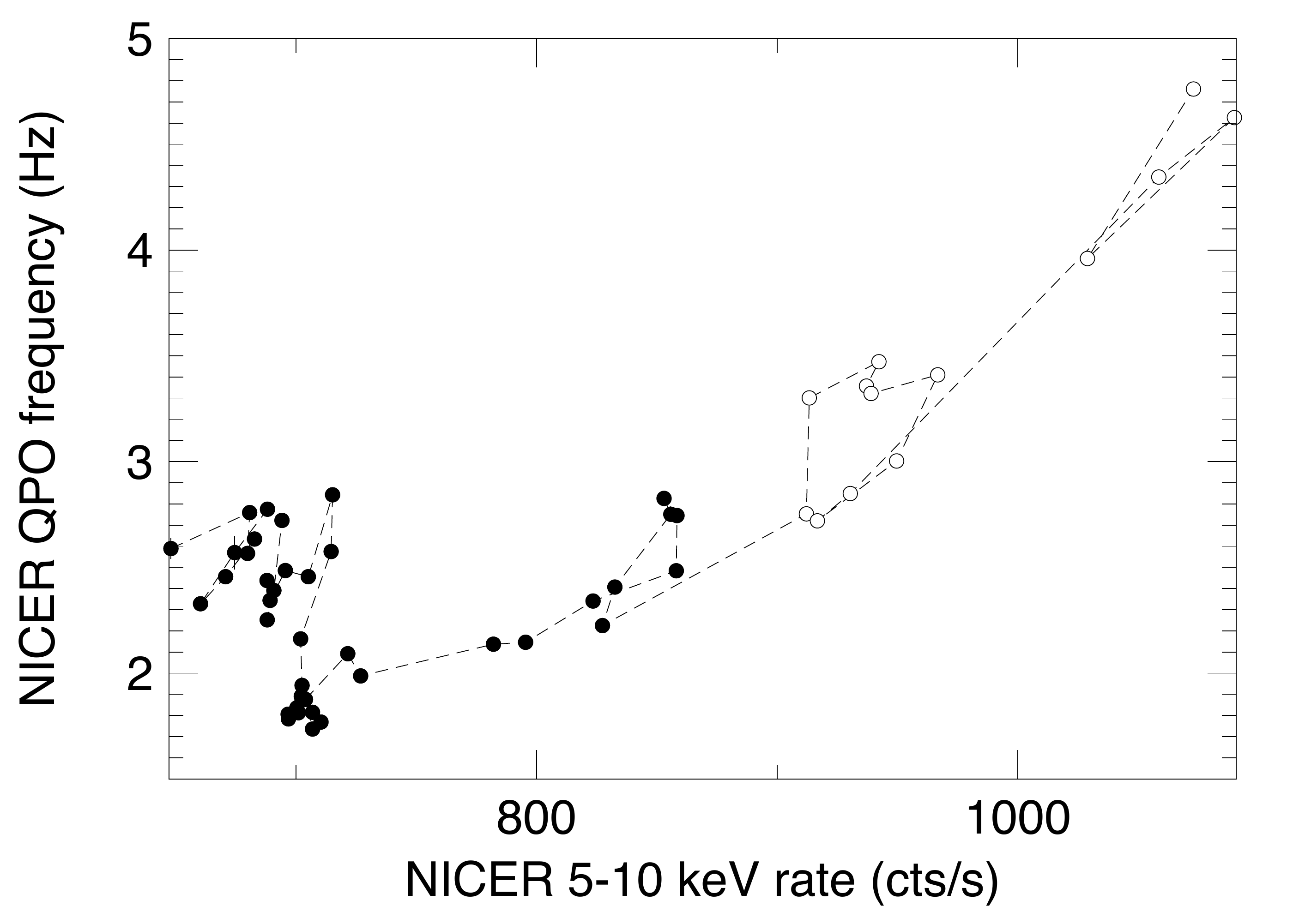} 
\caption{Correlation between QPO frequency and 5--10 keV count rate for the NICER data before MJD 58015 (before the peak of the outburst, see \autoref{fig:nicer_licu}). The black points indicate observations within the \astr\ observing window.}
\label{fig:nicer_rate}
\end{figure}

In order to put our results in a more general context of the outburst, we analysed the available NICER data of the rising part of the outburst of \src, corresponding to the points in \autoref{fig:nicer_licu} before MJD~58015 (up to just before the large increase in count rate).
The analysis was performed in the same way as described before for the \astr\ data using all available counts detected by NICER. The $\nu_{\rm{QPO}}$ was also very clear and easy to detect in the PDS despite the fact that the typical segment duration was $\sim$6 minutes.
In \autoref{fig:nicer_rate} we plot the NICER $\nu_{\rm{QPO}}$ as a function of the NICER 5--10~keV rate, chosen in order to be close to the laxpc energy band. The points corresponding to times overlapping with the \astr\ observing window are marked in black. One can see that the results are consistent for the overlapping period, with no correlation between frequency and count rate. However, after \astr\ stopped observing there is a clear positive correlation between the two parameters.
\yb{\citet{Mereminskiy2018AstL...44..378M} and \citet{Stiele2018ApJ...868...71S} also perform a spectro-timing analysis of this source and observe a similar positive correlation between the $\nu_{\rm{QPO}}$ and power law index. The extent of the observations analysed by these authors is larger than the \astr\ observation window and thus the tight correlation we observe is a subset of the correlation observed by them. }

\section{Discussion}

We have performed a spectro-timing analysis of the black hole binary \src\ using the observation conducted by \astr\ during the Hard Intermediate state of the source. The sources exhibits a secular roughly linear increase in the flux during the course of the observation. Fluctuations over the secular increase were observed in both softer and harder energy bands. The residuals from the linear fit for 3--10~keV and 30--80~keV are observed to be anticorrelated.

The energy spectrum of the source was modelled using an absorbed thermal disk blackbody with a power law as a non-thermal component. The high intrinsic absorption reported by \citep{Gendreau2017ATel10768....1G, Kennea2017ATel10731....1K, Xu2018ApJ...852L..34X} is also seen in current observation. The disk temperature remains typically around 0.2 keV with significant deviations in some segments. The secular increase in the total flux reflects an increase in both the flux of the non-thermal component and that of the thermal component (see \autoref{fig:para_var}). The power-law index of the hard component does not show a secular variation, but oscillates between 2.3 and 2.5, clearly being the cause of the anticorrelation in the residuals from the linear fits to the flux evolution in different bands. The oscillations are not random, but seem to follow a random walk around 2.4, with small variations between adjacent segments.

The PDS for individual segments show band-limited noise and a prominent QPO, with harmonics and subharmonics present in some of the segments. This shape of the PDS is typical of the HIMS  \citep[see][]{Belloni2011BASI...39..409B}.

The centroid frequency of the QPO as a function of time (see \autoref{fig:time_evolution}) oscillates between 1.7~Hz and 3.0~Hz and the oscillations follow those of the power-law index very closely (\autoref{fig:main_correlation}). \yb{The correlation for a larger observation window has been reported by \citet{Mereminskiy2018AstL...44..378M} and \citet{Stiele2018ApJ...868...71S} but due to sparser observations, the correlation is not as tight as seen in the present work. The trend observed by \citet{Mereminskiy2018AstL...44..378M} resembles the correlation observed for GRS~1915+105 \citep{Shaposhnikov2007ApJ...663..445S}}

Such correlation has been previously observed in several BHBs \citep{Vignarca2003A&A...397..729V} and provides a tight link between the QPO and the hard-component emission. In that work, the correlation was found on time scales of minutes in GRS 1915+105 and on scales of days-months for other transient BHBs. Here we follow the QPO over five days. \yb{\citet{Furst2016ApJ...828...34F} also observe an extremely tight correlation of the QPO centroid frequency and power law index in GX~339$-$4 in the frequency range of 0.6--1~Hz and $\Gamma$ of 1.66-1.8. The authors estimate the mass of the BH using a scaling relation provided by \citet{Shaposhnikov2007ApJ...663..445S}.   } 

The typical evolution of the outburst of a transient BHB is described best in the hardness-intensity diagram, where a counterclockwise path is followed. In the right ``vertical'' branch, corresponding to the LHS, the flux increases, the power-law index is well correlated with it and the frequencies of the timing features, although no peaked QPO is usually observed, also increase \citep[see][]{Motta2009MNRAS.400.1603M,Motta2011MNRAS.418.2292M}. Since the spectrum gradually softens, this branch is not really vertical. In the top ``horizontal'' branch the spectrum softens further, with the power-law index $\Gamma$ increasing in absolute value. The flux usually continues to increase, although not as fast as before, and the QPO frequency is correlated with both flux and $\Gamma$. Given the shape of the PDS and the values of the power-law index, our observations are clearly on the HIMS branch. As described before, usually both the LHS and the HIMS branch see a correlation between frequencies of variability components, the flux and the power-law index. In our \astr\ data we see the absence of a correlation with flux, although in the NICER data following our observation this correlation is present. This clearly shows that the QPO frequency is related to the energy spectral shape and not to the flux in either of the two main components. 
In a thermal-Comptonization scenario the index of the hard component steepens as the population of electrons cools due to the increase in soft photon input \citep[see e.g.][]{Motta2009MNRAS.400.1603M}. However, from  \autoref{fig:para_var} we can see that this is not the case here, as the secular increase in disk flux is not followed by $\Gamma$. Another parameter that determines $\Gamma$ is the optical depth of the cloud $\tau_0$: for a spherical cloud and input photon energy much lower than the temperature of the electron cloud $kT_e$ we have $\Gamma = -1/2+\sqrt{9/4+\gamma}$ where
$\gamma = \frac{\pi^2}{3}\frac{m_ec^2}{kT_e (\tau_0 + 2/3)^2}$ \citep{Sunyaev1979Natur.279..506S, Sunyaev1980A&A....86..121S}.
Therefore in order to keep $\Gamma$ from varying when the cloud temperature decreases is to increase $\tau_0$ in a very specific way. Fluctuations in this relation could in principle give rise to the observed fluctuations in $\Gamma$, but it would be very {\it ad hoc}.

Within the model that associates the low-frequency QPO with Lense-Thirring precession of the inner part of the accretion flow \citep[see][and references therein]{Ingram2011MNRAS.415.2323I, Ingram2016AN....337..385I} the QPO frequency is related to the size of the inner flow portion that precesses. With increasing accretion rate this region becomes smaller and the frequency increases. In our case, if the flux is a proxy for accretion rate, then this does not work. The same can be said of the Transition-Layer model \citep[see][]{Titarchuk2004ApJ...612..988T, Shaposhnikov2009ApJ...699..453S}, where also QPO frequency is related to accretion rate. \yb{\citet{Xu2017ApJ...851..103X} attribute the increase of the QPO frequency to the inward motion of the inner accretion edge, which they test by comparing the QPO frequency with the inner truncation radius obtained from the spectral fitting. The inner hot component being optically thin could explain the non thermal emission observed in \src. }

\yb{The correlation between the QPO frequency and the power law index is observed to be tighter than the correlation between the QPO frequency and the flux in the duration of the \astr\ observation. The correlation between QPO frequency and flux picks up subsequent to the \astr\ observation. The strong correlation we observe implies that the fluctuations manifesting as QPOs are closely related to the comptonising region as opposed to typical models assuming the origin of QPOs in the accretion disk. } 

In conclusion, the $\nu_{\rm{QPO}}$-$\Gamma$ relation, which is a common property of the hard states of BHBs (LHS and HIMS), appears here more complex than previously known and its origin must be investigated in detail, as none of the current models appears to be able to reproduce the results of this observation.

\section*{Acknowledgements}
This work makes use of data from the \astr\ mission of the Indian Space Research Organisation (ISRO), archived at Indian Space Science Data Centre (ISSDC). 
The authors would like to acknowledge the support from the LAXPC Payload Operation Center (POC) and SXT POC at the TIFR, Mumbai for providing support in data reduction. YB acknowledges the help of his colleagues in understanding the \astr\ data. 
This work has been supported by the Executive Programme for Scientific and Technological cooperation between the Italian Republic and the Republic of India for the years 2017-2019 under project IN17MO11 (INT/Italy/P-11/2016 (ER)). TMB acknowledges financial contribution from the agreement ASI-INAF n.2017-14-H.0. 





\bibliographystyle{mnras}
\bibliography{ref.bib} 


\begin{table*}
\caption{List of start and end points of the segments of the data. The time stamps are measured from MJD~58008.24759259 in seconds. The spectral and timing parameters with 90\% confidence interval for each of the segments are also tabulated. Due to jitter in the satellite, the pointing of SXT went off source for certain orbits due to which no SXT data was available for those segments. The spectral parameters could not be constrained for these segments and thus they have not been reported here. }
\label{tab:time_stamps}
\scriptsize
\begin{tabular}{ccccccc}
Segment number & Start time$^{\rm{a}}$ (in s)  & End Time$^{\rm{a}}$ (in s)   & Power law index   		& QPO frequency (in Hz) &  &  \\ \hline
1              & 0          & 3735.296   & ---                 		& $2.161\pm	0.006$ &  &  \\
2              & 3735.296   & 9777.92    & $2.380 \pm 0.011$        & $2.188\pm	0.004$ &  &  \\
3              & 9777.92    & 15622.784  & $2.416 \pm 0.011$     	& $2.406\pm	0.006$ &  &  \\
4              & 15622.784  & 21468.8    & $2.438 \pm 0.011$     	& $2.514\pm	0.007$ &  &  \\
5              & 21468.801  & 27314.048  & $2.450 \pm 0.011$    	& $2.597\pm	0.008$ &  &  \\
6              & 27314.049  & 33160.064  & $2.474 \pm 0.011$      	& $2.717\pm	0.006$ &  &  \\
7              & 33160.062  & 41366.911  & $2.468 \pm 0.011$     	& $2.773\pm	0.013$ &  &  \\
8              & 41366.91   & 47638.911  & $2.468 \pm 0.011$     	& $2.573\pm	0.009$ &  &  \\
9              & 47638.91   & 53910.911  & $2.423 \pm 0.011$   		& $2.486\pm	0.001$ &  &  \\
10             & 53910.91   & 60166.529  & $2.408 \pm 0.012$     	& $2.388\pm	0.006$ &  &  \\
11             & 60166.527  & 66405.759  &  ---        				& $2.753\pm	0.006$ &  &  \\
12             & 66405.758  & 72710.525  &  ---  					& $2.820\pm	0.006$ &  &  \\
13             & 72710.531  & 78982.525  & $2.449 \pm 0.013$      	& $2.600\pm	0.005$ &  &  \\
14             & 78982.531  & 85254.525  & $2.415 \pm 0.011$   		& $2.370\pm	0.006$ &  &  \\
15             & 85254.531  & 91526.915  & $2.384 \pm 0.013$     	& $2.272\pm	0.004$ &  &  \\
16             & 91526.914  & 97453.697  & $2.439 \pm 0.011$     	& $2.518\pm	0.004$ &  &  \\
17             & 97453.695  & 103283.329 & $2.453 \pm 0.011$  		& $2.561\pm	0.007$ &  &  \\
18             & 103283.328 & 109129.345 & $2.491 \pm 0.012$		& $2.771\pm	0.004$ &  &  \\
19             & 109129.344 & 114975.361 & $2.447 \pm 0.011$	    & $2.542\pm	0.006$ &  &  \\
20             & 114975.359 & 120823.806 & $2.432 \pm 0.011$		& $2.505\pm	0.005$ &  &  \\
21             & 120823.805 & 129158.915 & $2.400 \pm 0.011$		& $2.201\pm	0.010$ &  &  \\
22             & 129158.914 & 135430.915 & $2.357 \pm 0.011$		& $1.972\pm	0.008$ &  &  \\
23             & 135430.906 & 147942.525 & $2.313 \pm 0.011$		& $1.813\pm	0.007$ &  &  \\
24             & 147942.531 & 154230.915 & 	--- 					& $1.792\pm	0.005$ &  &  \\
25             & 154230.906 & 160502.915 &  ---						& $1.809\pm	0.005$ &  &  \\
26             & 160502.906 & 166774.915 & $2.332 \pm 0.012$		& $1.838\pm	0.005$ &  &  \\
27             & 166774.906 & 173046.915 & $2.325 \pm 0.011$		& $1.844\pm	0.005$ &  &  \\
28             & 173046.906 & 179269.759 & $2.305 \pm 0.011$		& $1.725\pm	0.004$ &  &  \\
29             & 179269.766 & 185114.618 & $2.312 \pm 0.011$		& $1.789\pm	0.004$ &  &  \\
30             & 185114.625 & 190944.634 & $2.359 \pm 0.011$		& $2.024\pm	0.004$ &  &  \\
31             & 190944.641 & 196807.04  & $2.360 \pm 0.011$		& $2.039\pm	0.005$ &  &  \\
32             & 196807.047 & 202653.056 & $2.344 \pm 0.011$		& $1.960\pm	0.006$ &  &  \\
33             & 202653.062 & 208485.884 & $2.349 \pm 0.011$		& $1.982\pm	0.005$ &  &  \\
34             & 208485.891 & 216951.29  & $2.385 \pm 0.011$		& $2.264\pm	0.012$ &  &  \\
35             & 216951.297 & 223223.29  & $2.425 \pm 0.011$		& $2.500\pm	0.007$ &  &  \\
36             & 223223.297 & 229446.15  & $2.446  \pm 0.011$		& $2.469\pm	0.010$ &  &  \\
37             & 229446.141 & 235718.15  & $2.405 \pm 0.014$		& $2.144\pm	0.007$ &  &  \\
38             & 235718.141 & 242023.29  &  ---    					& $2.381\pm	0.006$ &  &  \\
39             & 242023.297 & 248295.29  &  ---    		            & $2.124\pm	0.008$ &  &  \\
40             & 248295.297 & 254567.29  & $2.386 \pm 0.011$		& $2.129\pm	0.006$ &  &  \\
41             & 254567.297 & 260839.29  & $2.388 \pm 0.011$		& $2.113\pm	0.004$ &  &  \\
42             & 260839.297 & 266931.072 & $2.371 \pm 0.011$		& $2.039\pm	0.004$ &  &  \\
43             & 266931.062 & 272775.947 & $2.409  \pm 0.011$		& $2.251\pm	0.010$ &  &  \\
44             & 272775.938 & 278621.197 & $2.426 \pm 0.011$		& $2.311\pm	0.005$ &  &  \\
45             & 278621.188 & 284467.572 & $2.415 \pm 0.011$		& $2.290\pm	0.009$ &  &  \\
46             & 284467.594 & 290313.603 & $2.406 \pm 0.011$		& $2.252\pm	0.006$ &  &  \\
47             & 290313.594 & 296157.322 & $2.398 \pm 0.011$		& $2.235\pm	0.008$ &  &  \\
48             & 296157.312 & 298470.915 &  ---    		            & $2.261\pm	0.010$ &  &  \\
49             & 298470.906 & 304742.915 & $2.399 \pm 0.011$		& $2.150\pm	0.008$ &  &  \\
50             & 304742.906 & 311014.915 & $2.369 \pm 0.011$		& $2.005\pm	0.005$ &  &  \\
51             & 311014.906 & 317237.759 & $2.399 \pm 0.011$		& $2.167\pm	0.006$ &  &  \\
52             & 317237.75  & 323542.54  &  ---    					& $2.219\pm	0.008$ &  &  \\
53             & 323542.531 & 329814.54  &  ---    					& $2.178\pm	0.004$ &  &  \\
54             & 329814.531 & 336086.915 & $2.420 \pm 0.014$		& $2.284\pm	0.005$ &  &  \\
55             & 336086.906 & 342358.915 & $2.440 \pm 0.011$		& $2.373\pm	0.004$ &  &  \\
56             & 342358.906 & 348630.915 & $2.465 \pm 0.011$		& $2.531\pm	0.004$ &  &  \\
57             & 348630.906 & 354590.478 & $2.501 \pm 0.011$		& $2.755\pm	0.004$ &  &  \\
58             & 354590.469 & 360436.478 & $2.499  \pm 0.011$		& $2.740\pm	0.005$ &  &  \\
59             & 360436.469 & 366298.884 & $2.487 \pm 0.011$		& $2.657\pm	0.005$ &  &  \\
60             & 366298.875 & 372128.509 & $2.420 \pm 0.011$		& $2.254\pm	0.005$ &  &  \\
61             & 372128.5   & 377974.54  & $2.424 \pm 0.011$		& $2.302\pm	0.008$ &  &  \\
62             & 377974.531 & 386263.29  & $2.440 \pm 0.011$		& $2.355\pm	0.012$ &  &  \\
63             & 386263.281 & 392535.29  & $2.418 \pm 0.011$		& $2.292\pm	0.008$ &  &  \\
64             & 392535.281 & 405030.134 & $2.404  \pm 0.011$		& $2.103\pm	0.005$ &  &  \\
65             & 405030.156 & 411334.915 &  ---    					& $2.196\pm	0.006$ &  &  \\
66             & 411334.906 & 417623.29  &  ---   					& $2.077\pm	0.006$ &  & \\ \hline

\end{tabular}
\begin{flushleft}
$^{\rm{a}}$ Measured from MJD~58008.24759259
\end{flushleft}

\end{table*}

\bsp	
\label{lastpage}

\end{document}